\documentclass{article}

\pdfoutput=1
\usepackage{arxiv}

\usepackage[utf8]{inputenc} 
\usepackage[T1]{fontenc}    
\usepackage{hyperref}       
\usepackage{url}            
\usepackage{booktabs}       
\usepackage{amsfonts}       
\usepackage{nicefrac}       
\usepackage{microtype}      
\usepackage{lipsum}
\usepackage{graphicx}

\usepackage{amsmath} 

\newcommand{\VI}{\textit{VI}} 
        
\newcommand{\VItt}{$\textit{VI}(t,t') \,$}
\DeclareMathOperator*{\argmax}{arg\,max}
\DeclareMathOperator*{\argmin}{arg\,min}

\title{Data-driven unsupervised clustering of online learner behaviour}

\author{
  Robert L.~Peach \\
  Department of Mathematics \\
  Imperial College London\\
  London, SW7 2AZ \\
  \texttt{} \\
   \And
 Sophia N.~Yaliraki \\
  Department of Chemistry\\
  Imperial College London\\
  London, SW7 2AZ \\
  \texttt{} \\
     \And
 David Lefevre \\
  Imperial College Business School\\
  Imperial College London\\
  London, SW7 2AZ \\
  \texttt{} \\
     \And
 Mauricio Barahona \\
  Department of Mathematics\\
  Imperial College London\\
  London, SW7 2AZ \\
  \texttt{} \\
}

\begin{document} 




\maketitle


\begin{abstract}
The widespread adoption of online courses opens opportunities for analysing learner behaviour and optimising web-based learning adapted to observed usage. Here we introduce a mathematical framework for the analysis of time series of online learner engagement, which allows the identification of clusters of learners with similar online temporal behaviour directly from the raw data without prescribing \textit{a priori} subjective reference behaviours.
The method uses a dynamic time warping kernel to create a pairwise similarity between time series of learner actions, and combines it with an unsupervised multiscale graph clustering algorithm to identify groups of learners with similar temporal behaviour. To showcase our approach, we analyse task completion data from a cohort of
learners taking an online post-graduate degree at Imperial Business School.  Our analysis reveals clusters of learners with statistically distinct patterns of engagement, from distributed to massed learning, with different levels of regularity, adherence to pre-planned course structure and task completion. The approach also reveals outlier learners with highly sporadic behaviour. 
\textit{A posteriori} comparison against student performance shows that, whereas high performing learners are spread across clusters with diverse temporal engagement, low performers are located significantly in the massed learning cluster, and our unsupervised clustering identifies low performers more accurately than common machine learning classification methods trained on temporal statistics of the data.  Finally, we test the applicability of the method by analysing two additional datasets: a different cohort of the same course, and time series of different format from another university.

\end{abstract}

\section*{Introduction}

The application of data analytics to educational data has surged in the past few years facilitated by the adoption of online learning platforms \cite[]{VanBruggen2005}. However, in parallel to the increased access to detailed information,  it is crucial to identify both the right type of data and analytical approaches that will allow us to gain interpretable insights into online engagement and learning patterns~\cite[]{Lodge2017}. The process of learning extends over time and thus the analysis of temporal data is an important focus for educational data analytics. In this work, we describe a methodology for the study of time series data collected from the engagement of learners with the tasks and stages of online courses. The analysis of temporal statistics has been shown to provide a fruitful avenue to identify learners at risk of failure~\cite[]{Mahzoon2018}, predicting performance \cite[]{Papamitsiou2014}, dropping out of a course \cite[]{Ye2014,Ye2015,Taylor2014,Jiang2014}, or identifying learner behaviours \cite[]{Antonenko2012}. Despite such developments, a recent review of the field suggested that temporal analyses are currently insufficient in number, and that additional methodologies are required \cite[]{Wise2017}.

Temporal analytics has been used in the educational context to investigate \emph{massed} versus \emph{distributed} study modes, i.e., to compare the performance of learners that study the material `massed' (or `crammed') into a single study period to that of learners that `distribute' their study of the material across a number of shorter study periods. The general conclusion has been that distributed practice is the more effective strategy \cite[]{Gerbier2015}. The benefits of such `spacing effect' \cite[]{Ebbinghaus2013} have been shown over differing periods and within different contexts \cite[]{Toppino2014}, although other reports have noted that the effect does not apply to all learning contexts \cite[]{Donovan1999}. However, a feature of previous data analyses is that they generally allocate subjects in advance to one of the two pre-determined study modes. Indeed, pre-allocation is also an inherent restriction in \emph{supervised} machine learning approaches, where labels are assigned \textit{a priori} to train an algorithm. 

Recent studies have collected time series of learners' behaviours and used them to cluster learners according to pre-selected features of the data (e.g., task focus, resource usage, etc) chosen to describe different approaches to problem solving.  However, such methods are highly dependent both on the temporal features chosen as descriptors, which are based on specific knowledge of the data, as well as the number of groups that are obtained by the clustering. For example, a recent study extracted particular features from learners following a blended course (i.e., on two platforms: face-to-face and online) and identified four behavioural groups separated according to their differing levels of engagement across the two platforms~\cite[]{Carroll2017}. Such studies exemplify how the combination of temporal analytics and cluster analysis can provide insights of use to educators, course designers, and researchers in learning analytics \cite[]{Wise2017,Vwen2017}.

\begin{figure}[h]
    \centering
    \includegraphics[width=1\textwidth]{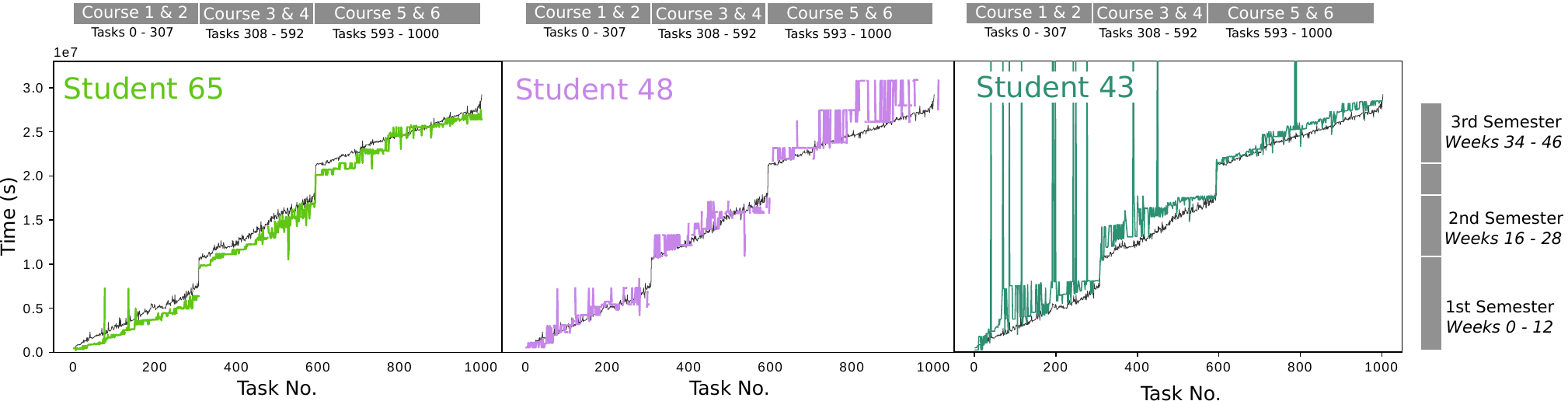}
    \caption{\textbf{The time-series of completion times for tasks are plotted for three example learners}. The entire course is divided into three semesters in which two modules are completed during each semester. The average task completion trajectory for the entire cohort of learners is plotted for comparison (black line). Learner 65 (left, light green) completes tasks consistently earlier than average, whereas learner 48 (middle, purple line) tends to finish tasks after the average, particularly in term 3. Learner 43 (right, dark green) exhibits rather sporadic behaviour during the first and second semesters whereby a lot of tasks are finished late or during massed learning sessions. The time series correspond to discrete events---the lines are guides to the eye conveying the temporal progression in task completion.}
    \label{fig:Fig1}
\end{figure}

Here, we present an unsupervised methodology that allows the direct analysis of raw time series gathered from the engagement of learners as they complete tasks of online courses without imposing \textit{a priori} neither the statistical descriptors of the time series nor the number or type of groups of learners to be detected. Hence the obtained learner clusters are not pre-determined or identified subjectively based on prior features but are detected algorithmically during the data analysis. To exemplify our approach, we analysed in detail the time series (i.e., time stamped data of task completion) of 81 learners as they undertook the six online compulsory courses that form the first year of a 2-year part-time postgraduate management degree. The courses extended over three terms and the patterns of task completion differ greatly across the learner group. Three examples of such highly distinct time series are shown in Figure~\ref{fig:Fig1}, showing a variety of behaviours: from steady completion to highly massed behaviour to sporadic patterns. 
To highlight its applicability, we also applied the method to two additional data sets: a different set of time series of task completion collected from the same degree programme but from a different year cohort, and a set of time series of online interactions (not of task completion) collected by a different university and therefore with distinct characteristics.

The methodology is summarised in Figure \ref{fig:Fig2}. We use the raw, time-stamped series of online actions from each learner and employ a dynamic time warping (DTW) kernel~\cite[]{Berndt1994} to calculate a similarity score between all pairs of learner time series.
Although several alternative methods exist to measure the similarity between two time series (e.g., Euclidean distance, Fourier coefficients, auto-regressive models, edit distance, or minimum jump models)~\cite[]{Serra2014},  DTW has been shown to outperform a variety of measures in classification tasks~\cite[]{MueenKeogh2016} and provides a principled way to use the full, raw information of the time series without preselecting features or functional representations~\cite[]{Wang2013b}. 
From the ensuing DTW similarity matrix, we construct a similarity graph, where nodes are learners and weighted links represent similarities between learners. This graph construction step is carried out
using the Relaxed Minimum Spanning Tree algorithm~\cite[]{Beguerisse-Diaz2013}, which aims to encapsulate the locally strong and globally relevant similarities in the dataset.  RMST has been shown to perform well in conjunction with the multiscale, unsupervised graph partitioning methodology of Markov Stability~\cite[]{Delvenne2010,Lambiotte2014}, which we apply to our graph to obtain clusters of learners with similar temporal behaviours. Alternative methods to cluster time-series data, with and without the creation of graphs, have been proposed in other contexts and applications~\cite[]{Rodrigues2008,Fenn2009,Ando2017,Hoffmann2018}. 
Instead of finding one particular clustering, our algorithm produces a multiscale description, given by a set of consistent clusterings of different coarseness obtained by robustly optimising
across all levels of resolution in an unsupervised manner, without pre-imposing the number or type of clusters (see Fig.~\ref{fig:Fig3}A for an example). 
Clusterings of different coarseness can then be used by the analyst according to their needs. If no robust clusterings are found, the algorithm will signal a lack of natural clusters in the data.
Details of the computational analysis are given in the Methods section. 

When applied to our case study dataset, our analysis identifies a set of clusterings of learners at different levels of resolution.  The clusters of learners reflect the differing temporal engagement as they progress through online course. In particular, our data-driven clusters capture behaviours associated with `massed'
(i.e., completion of a large number of tasks within a short time period)
 and `distributed' learning, as well as finer behaviours that differentiate these learning types into subgroups. 
For instance, at a coarse level, the algorithm identifies a cluster of learners that follow the course in a sequential and `distributed' manner; yet, at a finer resolution, this cluster is sub-divided into two clusters which differ by a 1-2 week difference in the average completion times of tasks (i.e., `early birds' and 'on time').  
Our approach also finds sporadic learners that skip a large number of tasks or exhibit irregular `massed' learning depending on particular courses or at different times of the year. Similar outcomes are observed for the other two datasets although with differences reflecting the particularities of the data.
We then used exam grades \textit{a posteriori} to establish whether  particular online engagement behaviours can negatively affect learner performance and we compared our groupings against classification based on statistical features computed from the time series data.

\section*{Results}

\subsection*{Unsupervised clustering reveals clusters of learners with differing online engagement}

\begin{figure*}[h]
    \centering
    \includegraphics[width=1\textwidth]{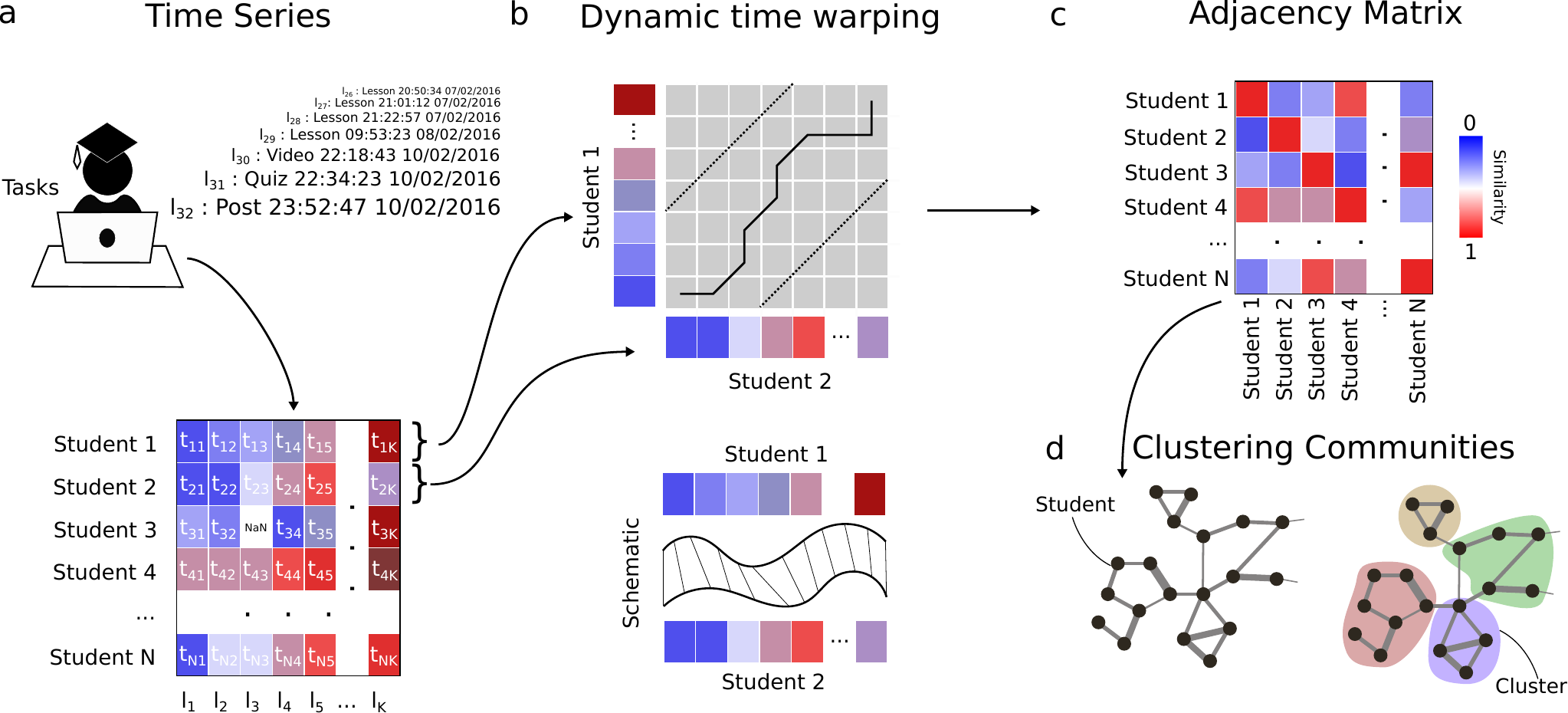}
\caption{ \textbf{A summary of the time-series clustering methodology} (\textit{A}) A time-series for each learner is compiled from their timestamps of interactions with tasks whilst undertaking the on-line course. (\textit{B}) Each learner time-series is compared pair-wise using a dynamic time warping kernel (see Methods) \cite[]{Berndt1994} which calculates a measure of similarity between two time-series. (\textit{C}) We construct an adjacency matrix $A$ where each element $A_{ij}$ describes the similarity between learners $i$ and $j$ pairwise. Similarity is measured between 0 and 1 (blue to red), where a higher value suggests a higher similarity. (\textit{D}) The adjacency matrix $A$ encodes a network where each represents a learner and the edges are the similarity between each learner. Community detection can be applied to cluster learners according to their similar time-series behaviours.}
    \label{fig:Fig2}
\end{figure*}

To find groups of learners with similar online engagement in an unsupervised manner, we follow the procedure summarised in Figure~\ref{fig:Fig2}. We first create a similarity matrix between learners using a Dynamic Time Warping kernel. This matrix is transformed into a similarity graph using a sparsification based on the Relaxed Minimum Spanning Tree~\cite[]{Beguerisse-Diaz2013}, a procedure that retains global network connectivity whilst discarding weak similarities that can be explained through longer chains of strong similarities. Through this process, we create a graph where the nodes are learners linked by edges weighted according to their time-course similarity. Hence, two learners that complete the tasks of the course in a similar manner will be linked by a strong edge. 

The constructed similarity graph is then analysed using Markov Stability (MS), a multiscale graph partitioning algorithm that uses a Markov process to scan the graph across Markov time in order to find optimised and robust partitions of the graph at any level of resolution~\cite[]{Delvenne2010, Lambiotte2014}. The partitions are found by maximising a resolution-dependent cost function (the Markov Stability) at all levels of resolution, as given by the Markov time, $t$. We then select robust partitions in the following sense: (i) they are persistent across scales (i.e., optimal over an extended Markov time $t$, as given by a plateau with a low value of $\text{VI}(t,t')$), and (ii) robust to the small changes in the optimisation (i.e., consistently found as a good partition over those scales, as given by a relative dip in $\text{VI}(t)$).
Such robust partitions identify clusters of learners that exhibit similar online temporal patterns. The definitions
of the different measures and some details of the Markov Stability framework are given in Methods.

Figure~\ref{fig:Fig3}A summarises the results of our multiscale clustering method applied to the time series of task completion of six online courses by 81 learners pursuing a post-graduate part-time Management degree at Imperial College Business School over one year. See Methods for further details about the data. 
As the Markov time is increased, the level of resolution is decreased and the method reveals robust partitions of decreasing granularity. In Figure~\ref{fig:Fig3}A, we illustrate the partitions found from 10 clusters down to 2 clusters, with a notably robust partition into 6 clusters.
Note the quasi-hierarchical aggregation of the finer clusters into coarser ones, a feature that is intrinsic to the data and not imposed by our clustering algorithm. 
(For a more detailed view of the multiscale clustering structure, see Supplementary Figure~1.)
The quasi-hierarchical organisation across levels of resolution reflects the fact that subtle temporal details characterise the finer clusters, but broader similarities of the time profiles define the coarser clusters. 
Hence, our computational framework allows for adjustable granularity, which can be tailored to the needs of the analyst.

To exemplify the characterisation of the results in our dataset, we focus mainly on the 6-cluster partition, which contains four large groups and two single learners that remain unclustered due to their highly individual sporadic behaviour. The 6 cluster partition exhibits the largest relative drop in $\text{VI}(t)$ and a long plateau in $\text{VI}(t,t')$. The 10-cluster and 8-cluster partitions are equally of interest and provide a more refined clustering consistent with the 6-way partition, as seen in Figure~\ref{fig:Fig3}A. The coarser 2-cluster partition is also of interest: the two clusters are found to separate learners that exhibi distributed and massed learning. 
In the rest of the paper, we concentrate on a more detailed description of behaviours emerging from the 6 cluster partition, as it provides a nuanced, data-driven level of resolution on the data.

\subsection*{Characterisation of the clusters of online learners}

 \begin{figure*}[h!]
    \centering
    \includegraphics[width=0.78\textwidth]{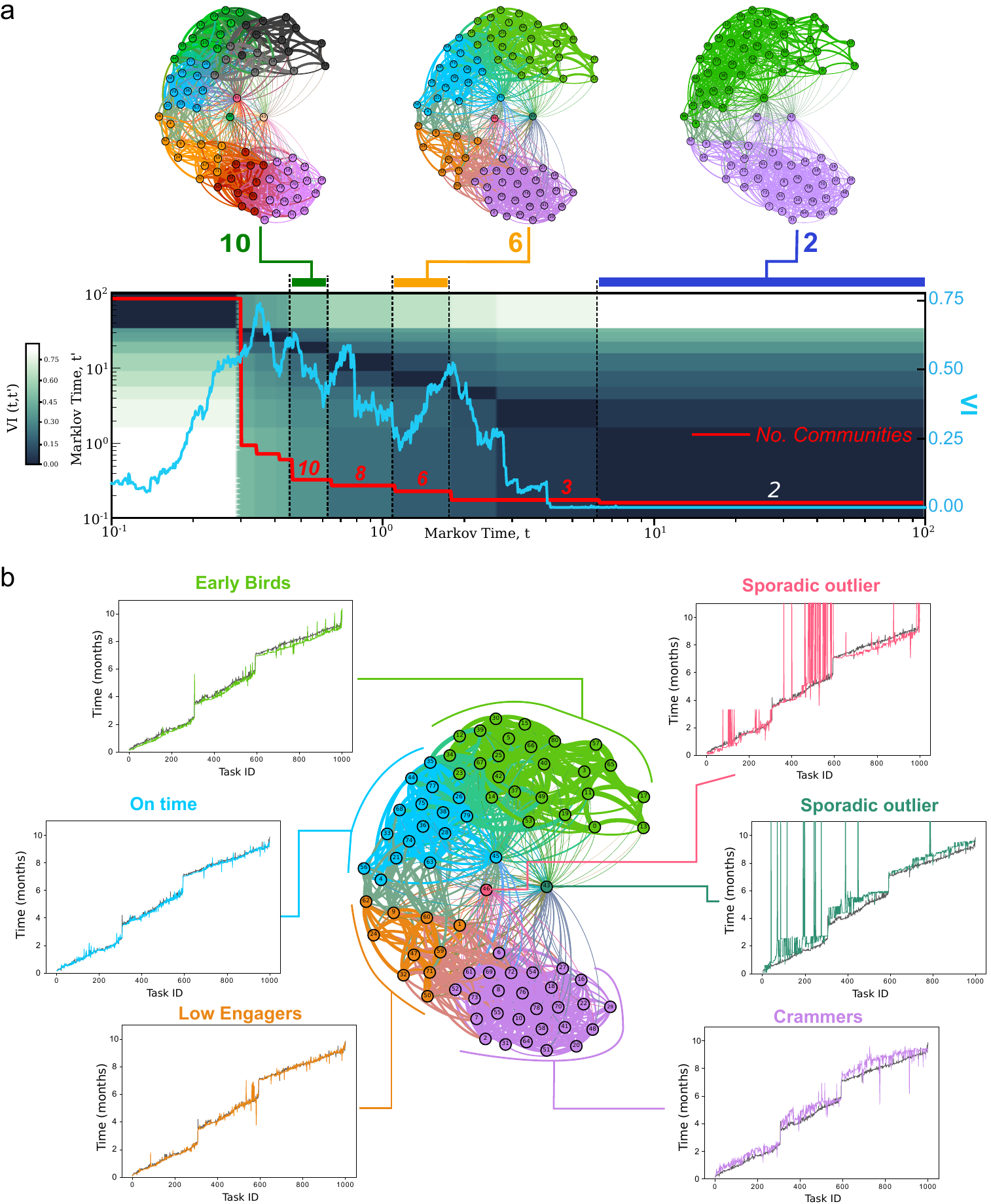}
 \caption{\textbf{Multiscale partitioning and characterisation of learner clusters.} (\textit{A}) Multiscale unsupervised partitioning of the similarity graph of learners using Markov Stability (MS). The nodes of the graph are learners ($N=81$) and edges between nodes are weighted according to the Dynamic Time Warping similarity~\ref{eq:DTW_sim} between temporal task completions. We obtain a quasi-hierarchy of graph partitions of increasing coarseness (number of clusters indicated by red line) which optimise the MS cost function~\ref{markovstability} over different spans of Markov time $t$. The learners within a cluster in the partitions exhibit similar online temporal behaviours at different levels of resolution.   
We seek partitions that optimise MS~\ref{markovstability} and are robust in two ways: consistent under the optimisation and persistent across scales. These are evaluated with the Variation of Information (\VI), a normalised metric of distance between clusterings akin to Mutual Information~\cite[]{Meila2003}. A low value (or dip) of $\VI(t)$~\ref{eq:VIt} (cyan line)  signals a partition that is consistently found by 100 repeated runs of the greedy Louvain algorithm.  A long plateau block (dark grey in the heatmap) of the $\text{VI}(t,t')$~\ref{eq:VItt} indicates a persistent partition across scales. Robust partitions according to both criteria include the10, 6 and 2-way partitions. See Supplementary Figure 4 for the graph layouts of the partitions, including the 8 and 3-way partitions.     
 (\textit{B}) For each cluster in the 6-way partition we display the average GPR trajectory for leaners in the cluster and the average GPR trajectory of the full cohort (black line).  
      Each community is given a descriptive tag (e.g., `Early Birds') following post-hoc evaluation of the average time series.}
    \label{fig:Fig3}
\end{figure*}

As shown in Figure \ref{fig:Fig3}B, the 6-cluster partition is both robust and the data-driven groupings it provides have an appropriate level of resolution to gain meaningful insight into the observed patterns of online learners.  
Two of the clusters contain only one learner, with highly individual and sporadic behaviour. For each of the other four clusters, we use Gaussian Process Regression (GPR) \cite[]{Rasmussen2010} to compute the average engagement trajectory of the group of learners, and compare it with the average GPR trajectory for the whole set of 81 learners. 
The computed GPRs allow us to quantify statistically the differences in the temporal patterns of the different clusters
using Bayes factors of the processes. In particular, we found that the trajectories of each cluster are statistically more probable to be derived from separate processes defined within their own cluster as follows.
A GPR was fitted to the entire set of trajectories and the log likelihood of the entire set of trajectories was calculated. Equally, the log likelihood of each separate cluster of trajectories from that same Gaussian Process was calculated. The Bayes Factor, calculated as the sum of log likelihoods of each separate cluster minus the log likelihood of the entire set of trajectories~\ref{bayesfactor} was found to be large (K = $3.37\times10^{10}$). This indicates that the behaviours of each cluster are statistically different from each other and are derived from different behavioural processes. This computation was repeated for the differences between each pair of neighbouring clusters. The Bayes factors were: $K=0.38\times10^{10}$ between the `early birds' and `on time' clusters; $K=1.52\times10^{10}$ between the `on time' and `low engager' clusters; and $K=0.17\times10^{10}$ between the `low engager' and `crammers' clusters.  These numbers provide statistical evidence of the differences between the obtained clusters.

Each of the clusters in this partition has been given a descriptive title that encapsulates the group behaviour. 
The learners in the `Early Bird' group  (green cluster)  generally exhibit a highly sequential and ordered approach to their learning and tend to complete their tasks earlier than the cohort average with a systematic 1-2 week advance offset.
The behaviour of learners in the 'On time' group (cyan cluster) is similar to the 'Early birds', except that they finish tasks closer to the average. Hence both the green and cyan groups present a similar `distributed learning' behaviour only distinguished by a slight delay, which explains why both groups are agglomerated into a single cluster in the coarser 2-way partition (Figure \ref{fig:Fig3}A). 
The learners in the `Low engagers' (orange) cluster also exhibit relatively distributed work flow (similar to the cyan and green clusters) but with less anticipation in the second half of the year (and especially in the third term). Furthermore, this group had a high number of tasks that were never completed.
The `Crammers' cluster (magenta) contained learners that exhibited massed learning (indicated by the presence of plateaux in their time-series, suggesting tasks being completed in a short period of time), low task completion and an ordering of task completion that deviates from the proposed course sequence. Finally, the outliers (learners 43 and 46), which form their own clusters, exhibit highly sporadic learning behaviours, with tasks completed at later dates without following sequentially the layout order of the course.

\begin{figure}[h]
    \centering
    \includegraphics[width=0.9\textwidth]{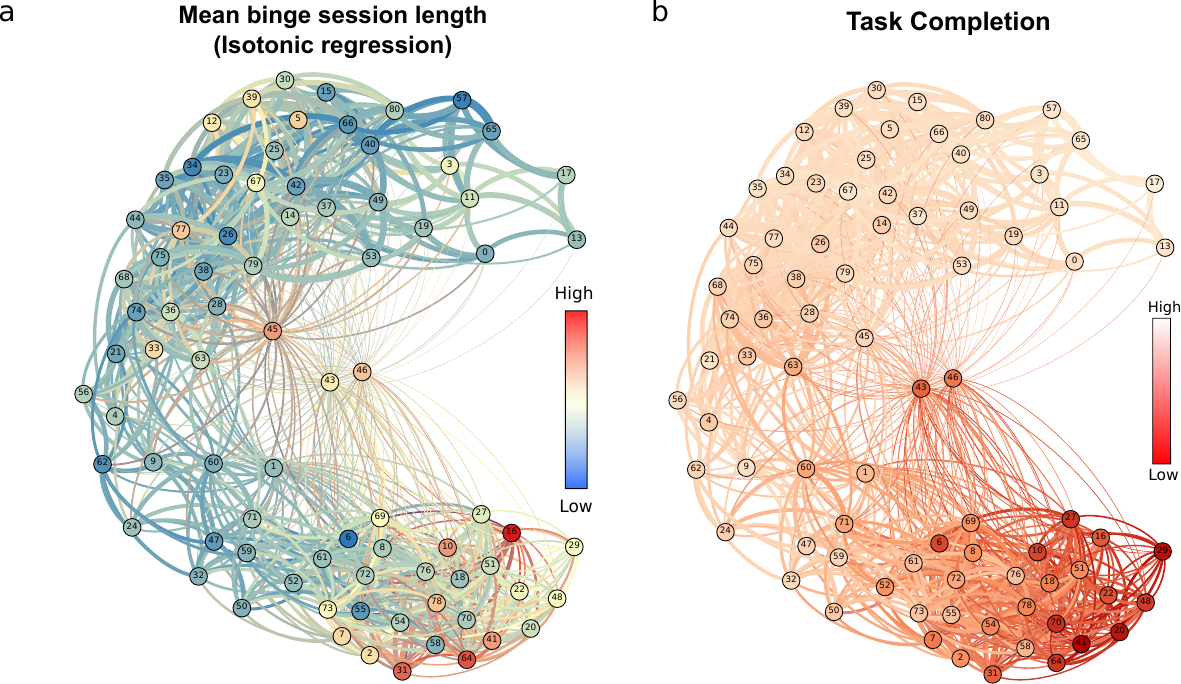}
\caption{\textbf{Deriving standard time-series metrics for individual learners.}
      (\textit{A})  The graph of learners with nodes (learners) coloured according to the mean learning session length. Colourmap extends from red (a lot of tasks completed within a small time frame) to blue (few tasks completed during each learning session). 
      (\textit{B}) The number of tasks completed by each learner is mapped onto the graph of learners.}      
    \label{fig:Fig4}
\end{figure}

To further characterise our results, we computed standard time-series metrics for each learner. Figure \ref{fig:Fig4} shows the graph of learners coloured according to two such statistical metrics derived directly from the time-series: the mean `massed' session length (commonly known as binge learning), and the percentage of completed tasks.
Figure \ref{fig:Fig4}A shows the mean massed session length, i.e., the length of plateau in the number of tasks over time calculated via an isotonic regression (see Methods). This measure captures events where a learner has completed a large number of tasks within a short time frame. We find that the `Crammers' cluster has a higher mean massed session length. 
Figure \ref{fig:Fig4}B shows the graph of learners coloured according to the percentage of tasks completed relative to the total number of available tasks. In general, the 'Crammers' cluster shows the lowest mean task completion (66\%), followed by a completion ratio of  80\% in the 'Low Engagers' group, and a higher mean task completion rates in the `On time' (86\%) and `Early Birds' (90\%) clusters. 

\subsection*{Cluster analysis identifies groups of learners at risk of low performance}

\begin{figure*}[h]
    \centering
    \includegraphics[width=0.9\textwidth]{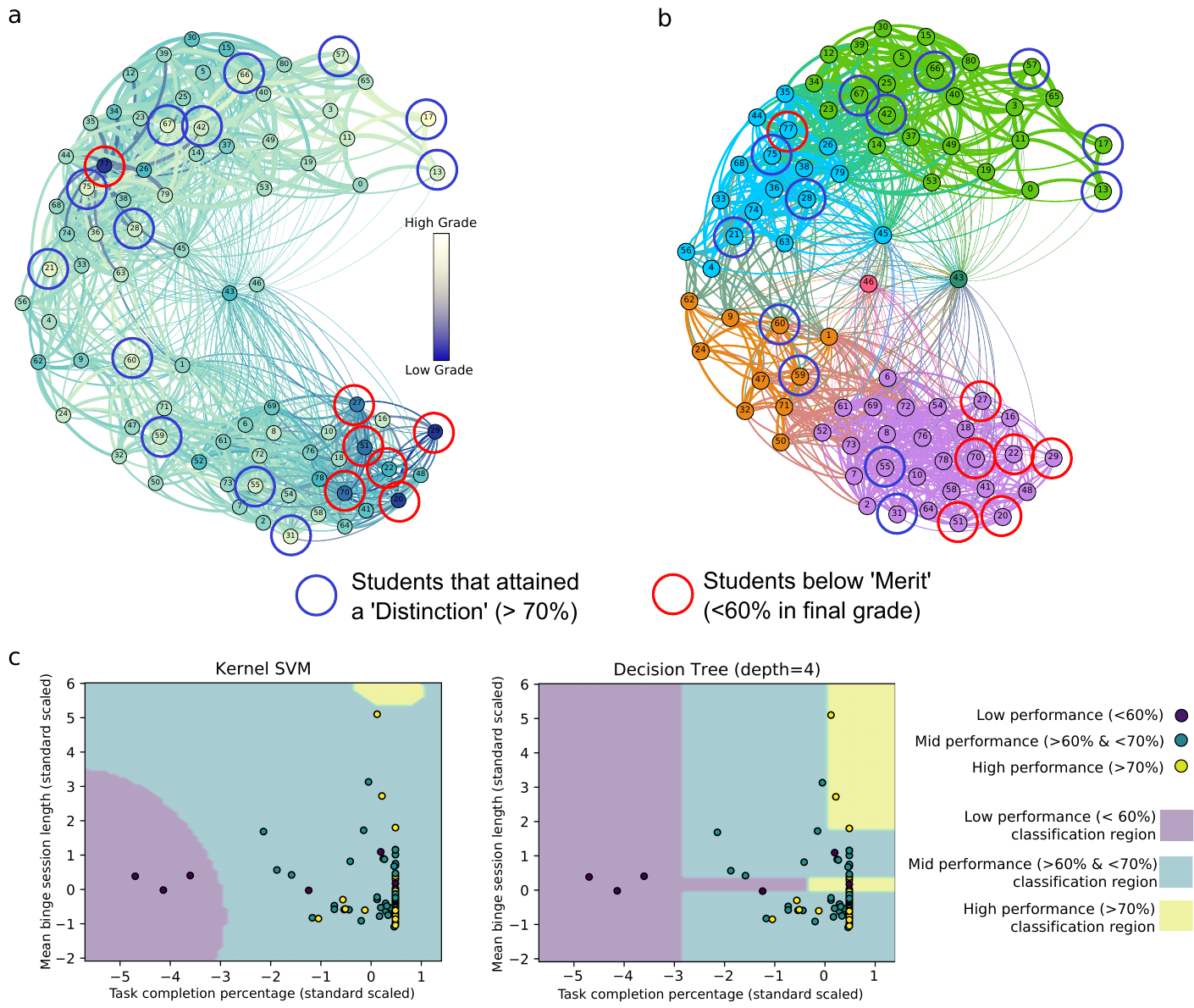}
\caption{\textbf{Evaluation of learner performance with respect to behavioural clusters. }          
      (\textit{A}) \textit{A posteriori} comparison of the clustering with the obtained marks shows that 6/7 learners that attained lower than 60$\%$ (red circles) concentrate in the `Crammers' massed learning cluster, whilst those that attained over $>$70$\%$ (blue circles) and the mid-range performers are distributed across all the clusters, with a higher concentration in the early bird distributed learning cluster.  A single low-performing outlier (learner 77) attained a low grade, yet displayed regular online patterns distinct from the other 6/7 low performing learners.
      (\textit{B}) The grades are mapped directly onto each learner in the network structure.
      (\textit{C}) Classification of learners using an SVM (RBF kernel) and a Decision Tree (depth 4) trained on two statistical measures from the time series: the mean binge session length and the percentage task completion (both normalised to a z-score, i.e., zero mean and unit variance).  The results show poor classification accuracy: only 3/7 low performance learners are accurately classified using the SVM and only 4/7 using the decision tree.}
    \label{fig:Fig5}
\end{figure*}

We have also carried out an \textit{a posteriori} evaluation of our behavioural clusters with respect to the performance of the learners. 
Figure \ref{fig:Fig5}A  shows the mapping of the final average marks on the learner graph, where we have also highlighted high performing ($>$70\%, top 15\%, `Distinction') and low performing ($<$60\%, bottom 7.5\%, below `Merit') learners.
Figure~\ref{fig:Fig5}B shows that  6 out of 7 low performance learners lie in the `Crammers' cluster associated with massed learning and reduced task completion. There was a specific learner (77, cyan cluster) who attained a low grade and yet did not exhibit time-series behaviours indicative of a low performance. 
The high performers tend to be distributed across all other clusters, suggesting that the learning behaviours of a high performer are not as critical to their success. Still, 9 out of the 13 high performing learners  are found in the 'Early Birds' or 'On time' clusters characterised by a sequential approach to their learning with minimal massed learning sessions.  
The sporadic learners in single clusters (43 and 46) did not attain either a low performance or a distinctly high one.

Although our method captures information congruent with time-series statistical metrics (e.g., those shown in Figure \ref{fig:Fig4} related to massed learning and task completion rates), the data-driven clusters we obtain encompass global time-series information beyond such pre-determined standard statistical measures.
To test this idea, we compared the results of our data-driven clusters to standard classification methods from Machine Learning based on statistical features.  
Figure \ref{fig:Fig5}C illustrates the classification map obtained by training two common machine learning algorithms using the two statistical features in Figure \ref{fig:Fig4}. The first learning algorithm is a support vector machine (SVM) using a radial basis function kernel and the second is a decision tree with a depth of 4 branches \cite[]{PedregosaFabian2011} (see Methods).
For both methods, we find that the accuracy of learner classification against performance is low: only 3-4 out of 7 low performance learners were accurately predicted.
This result suggests that using a finite set of pre-determined time-series features reduces the information available to differentiate the necessary behaviours relevant to performance. In contrast, our graph construction and clustering methodology utilises the full content of the time series (including attributes that are not evident from inspection of particular statistical metrics), thus providing a more comprehensive  grouping of learners with similar temporal behaviours.

\subsection*{Testing the methodology on two additional datasets}

\begin{figure*}[h]
    \centering
    \includegraphics[width=0.75\textwidth]{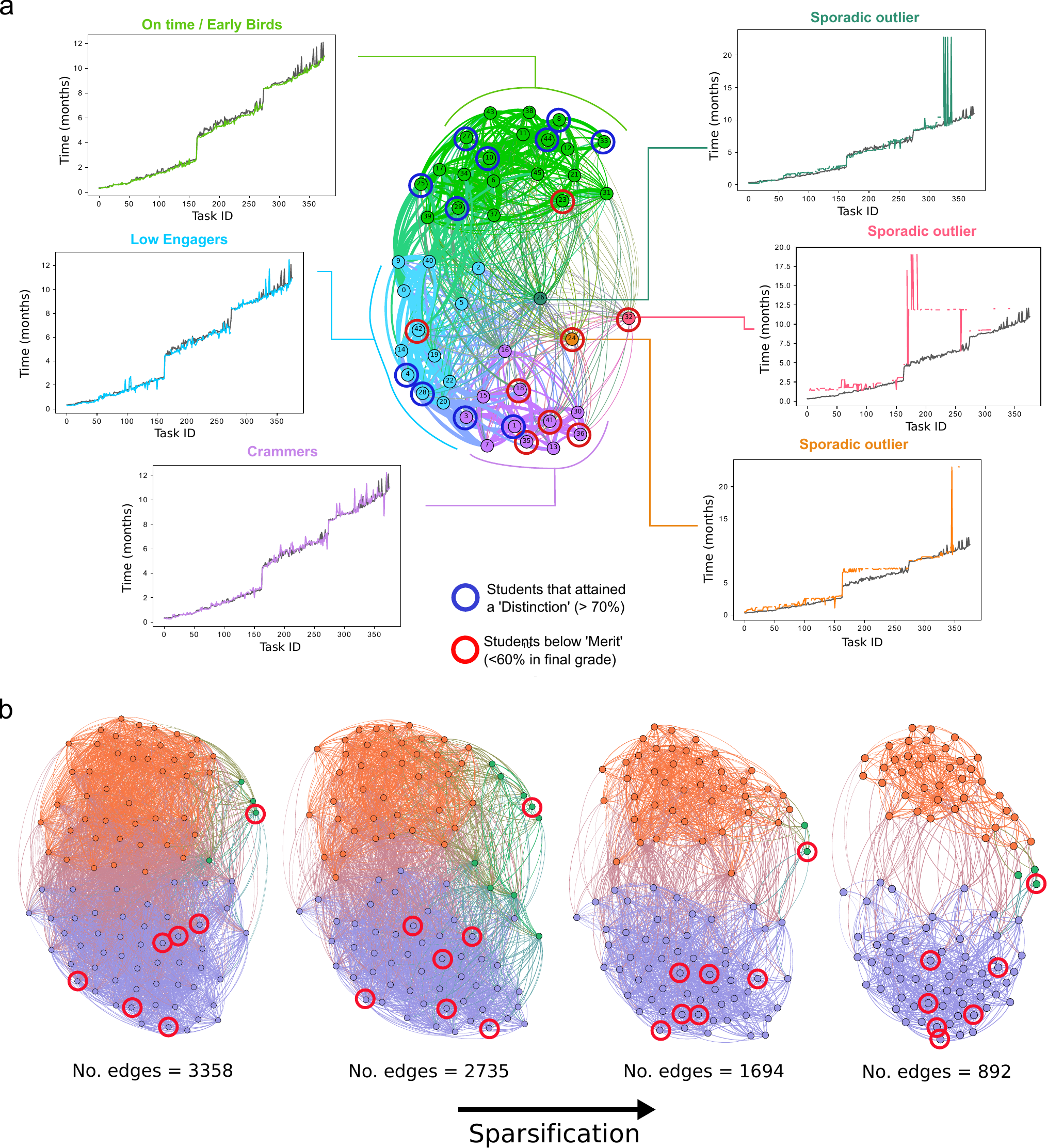}
\caption{\textbf{Validation of the methodology against two additional datasets. }          
      (\textit{A}) Application of our methodology to a second cohort of learners ($N=46$) undertaking a similar online management course at Imperial Business School with a different set of tasks (No. tasks = 376) over a 1-year period. Temporal behaviours associated with the 6-way partition revealed similar behaviours to the cohort in Fig.~\protect{\ref{fig:Fig3}}. We also find that 6/8 of the low performance students (red circles) are found in the `Crammers' cluster or displayed sporadic engagement behaviour, and only 1/8 is found in the green cluster of distributed learning behaviours. The high performers, on the other hand, are diverse in their patterns of online engagement, as also observed in the first cohort. For the full analysis of this second cohort see Supplementary Figure 2A. 
      (\textit{B}) Analysis of  a cohort of students undertaking an anonymised course at the Open University (OULAD dataset) over 250 days \cite[]{Kuzilek2017}. In this case, the time-stamped data corresponds to `page clicks' rather than task completions. The Markov Stability analysis finds a robust 3-way partition consistently across all level of sparsifications of the dataset (full analysis in Supplementary Figure 4). Within the 3-way partition, there are two major clusters (purple and orange) and one minor cluster (green). The purple cluster corresponds to massed learning and contains 6/7 of the low performers with a significantly lower grade (paired t-test at 5\%) relative to the orange cluster of distributed learning behaviours. This difference exists at all levels of network sparsification.
The minor cluster (green) includes outlier behaviours. }
    \label{fig:Fig6}
\end{figure*}

We have applied the methodology to analyse task completion time series data from a second cohort of 46 learners taking the online management course at Imperial College Business School. The results we obtain are similar, as shown in the multiscale clustering presented in Supplementary Figure~2 and the detailed analysis of the 6-cluster partition in Fig.~\ref{fig:Fig6}A. In this case, we identified a robust 9-cluster partition (with 4 major clusters and 5 single learner clusters) and a robust 6-cluster partition (with 3 major clusters and 3 single outliers). The major clusters in the 6-way partition (shown in Fig.~\ref{fig:Fig6}A) showed similar behaviours to those observed in the first cohort we analysed. In particular, the green cluster in Fig.~\ref{fig:Fig6}A  corresponds to the `Early Birds' and `On time' groups in Fig.~\ref{fig:Fig3}, whereas the blue cluster in Fig.~\ref{fig:Fig6}A is similar to the group of task-skipping  `Low Engagers' group in Fig.~\ref{fig:Fig3}, and the purple cluster in Fig.~\ref{fig:Fig6}A exhibits similar traits to the 'Crammers' cluster in Fig.~\ref{fig:Fig3}.  
Within this 6 cluster partition, we found that of the 8 low performance learners, 4/8 were located in the massed learning cluster, 2/8 were sporadic outliers, and 1/8 was in the low engagement cluster. Only 1/8 was located in the distributed learning cluster. Moreover, using standard classification procedures in Supplementary Figure~3 we found that our methodology was superior at grouping learners with similar performance. These findings highlight the consistency of the methodology across the cohorts, yet attuned to particularities of the data.

The types of temporal engagement data collected from learners will differ across educators or institutions depending on the particularities of the Learning Management System. To test the methodology on a different kind of data, we have studied a set of 100 learners undertaking an anonymised course within the Open University (OULAD dataset \cite[]{Kuzilek2017}). The OULAD dataset differs from our dataset in several ways: (i) the time-stamp data in OULAD corresponds to page clicks and not necessarily to task completion; (ii) the time stamps were coarse-grained to days; (iii) pages could be revisited. 
The results of applying our methodology to the OULAD dataset in Fig.~\ref{fig:Fig6}B (and Supplementary Figure~4 of the Supplementary Information) show that the multiscale clustering is robust to the sparsification implicit in the graph creation step.
A robust 3-way partition is consistently found in our analysis, with two major clusters and a minor cluster of outliers. The two major clusters corresponded to a separation of learners who exhibited higher massed learning and lower task engagement versus learners with a distributed learning. We found that 6/7 of the low performance learners ($< 60\%$) were located in the cluster associated with massed learning, whilst one low performance learner was located in the minor outlier cluster and none were in the distributed learning group.

\section*{Discussion}

We have described an approach for the analysis of temporal data of online learning behaviours, in which distinct clusters of learners are obtained algorithmically without using \textit{a priori} statistical information about individual behaviours or about the number or type of expected behaviours across the cohort. The mathematical framework is general, and can be applied broadly to any time series data in physical or social sciences to identify distinct temporal behaviours. 
In the context of learning analytics, we showcased the method through three data sets of online learner activity of different types and origins.

Our method uses a dynamic time warping similarity kernel to generate a sparsified similarity graph between learners, to which we apply a multiscale graph partitioning algorithm in order to find optimised and robust clusters of learners with similar temporal behaviours at any level of resolution in an unsupervised manner. Because our method uses the full time series, it inherently encompasses richer temporal information than standard methods based on selecting statistical features of the time series.

In the data sets analysed here, we obtained a quasi-hierarchy of robust partitions, from finer to coarser, which provide different levels of information, as required by the analyst. 
For instance, in our main case study in Figure~\ref{fig:Fig3}, we found robust partitions into 10, 6 and 2 clusters. The 6-way partition consists of four large learner clusters (`Early Birds', `On time', `Low Engagers' and `Crammers')  and two single unclustered learners (`Sporadic outliers'), which were shown to be statistically different to each other according to the GPR Bayes factor~\ref{bayesfactor}. 
\textit{A posteriori} comparison with learner performance indicates good correspondence with the obtained clustering:  low performers are generally located (6 out of 7) in the `Crammers' cluster (associated with massed learning and low task completion) and are generally absent from the `Early Birds' and `On time' clusters (associated with distributed learning and high task completion).  On the other hand, high performers are distributed across several clusters, albeit with higher prevalence in the clusters associated with distributed learning. 
These results provide an improved characterisation as compared to common machine learning classification algorithms trained on two statistical measures from the time series. 
The analysis could be enhanced by the use of finer partitions (e.g., the 10-way partition has clusters with over-representation of low performers (purple cluster, hypergeometric p-value = 0.00023) and high performers (charcoal cluster, hypergeometric p-value = 0.026), as seen in Supplementary Figure~1). 
Similar general behaviours and classifications are obtained for the two additional data sets presented in Fig~\ref{fig:Fig6} and the Supplementary Information.
 
The fact that low performers tend to concentrate in the massed learning cluster and be absent from the distributed learning clusters is in agreement with previous studies which found that learners that `crammed' retained less information when tested at a later date \cite[]{Bloom1981}, and provides support for the risks associated with this behaviour. 
On the other hand, the fact that high performers are distributed across several clusters, 
albeit with higher prevalence in the clusters with high task completion and distributed learning, 
suggests they follow a host of diverse learning patterns, in agreement with a latent class model that suggested that the `spacing effect' is less prominent for high performers~\cite[]{Verkoeijen2008}. 
These observations were found to be consistent when testing our methodology on a second cohort of learners within the same institution and online degree, and broadly in agreement with a different type of data ('page clicks') from a set of online learners at the Open University, where we found a strong distinction between a low performing `massed learning' cluster \textit{vs.} a `distributed learning' cluster.

Clearly, temporal behaviours do not fully account for learner performance, and this methodology is not intended as a diagnostic tool, but rather as providing a method to explore and identify learner engagement behaviours with the purpose of aid, intervention and help with course design. Combining the temporal analysis introduced here with established `early warning system' analyses \cite[]{Beck2001} could aid in such tasks. Although educators might encourage learners to pursue a distributed study behaviour, our results suggest a nuanced approach for high performers, with flexibility provided in course design so that high performing learners may pursue the study strategies they personally find effective.

Future work within different learning contexts, coupled with additional dependent variables of interest (e.g., learner satisfaction, career success, interruption and withdrawal rates) could be important to provide broader support for the initial results reported here. We remark that the methodology is scalable to larger datasets through adjustments of the computation of both the DTW kernel and the Markov Stability cost function (see Methods).  Further improvements of the similarity kernel using constrained DTW \cite[]{Ratanamahatana2013} and end point invariance \cite[]{Silva2016} could also be used to improve the sensitivity and accuracy of the method in representing the different temporal behaviours. Together with how online behaviour changes over time for each of the learners, these directions will constitute areas of further research.

\section*{Methods}

The methods section describes the data and unsupervised mathematical pipeline used to analyse the trajectories of learners. The research was performed without any \textit{a priori} knowledge or allocation of the learners, making it similar to a blind investigation. 

\subsection*{Temporal Data}

The main case study of this research was based on task completion data from 81 post-experience learners pursuing a post-graduate part-time management degree at Imperial College London. These learners formed part of a cohort of 87 learners. Data from the remaining 6 learners was not included here as these learners either interrupted their studies or withdrew from the programme. Subjects ranged in age from 28 to 53 years old, with gender balance of 57 males to 24 females, and they resided in 18 geographically disparate countries. The data corresponds to interactions with six online courses which together comprised the first academic year of the 2-year degree programme.  Although the subjects met face-to-face at the start of each academic year, the six courses were studied completely online. Subjects proceeded in a lock-step manner through the  academic year which  was split into three 10-week terms each containing  two of the six courses. The anticipated study load was  5 to 7 hours per week for each course, so 10 to 12 hours in total. The courses were assessed via a combination of coursework and exam, however, participation in these separate assessed activities was not included in the dataset analysed here, only their final 2-year grade was used as an indication of their performance.

To highlight the applicability of the method, we also applied the analysis to two additional datasets: 
(i) time stamped task completion series from a second cohort of 46 post-experience learners pursuing the post-graduate part-time management degree at Imperial College London; 
(ii) time stamped data of 'page-clicks' (not equal to task completion) from 100 learners undertaking Open University courses (OULAD dataset~\cite[]{Kuzilek2017}).  
For further details on these data sets, see the Supplementary Information.

Ethical approval from the Education Ethics Review Process (EERP) at Imperial College London was attained (EERP 1718-032b) and a waiver for informed consent was granted for this study.

\subsection*{Construction of the learner similarity graph using a Dynamic Time Warping kernel and RMST sparsification.}

\subsubsection*{Creating a similarity matrix between learners using Dynamic Time Warping}
To compute the similarity between the task completion time traces of every two learners $i$ and $j$, we use a similarity kernel, i.e., a generalised inner product. Common approaches for sequence analysis use $L_p$ norms (when $p=2$ we obtain the Euclidean norm) which are fast to compute and easy to index. However, their one-to-one matching often ignore sequential patterns that are non-linearly misaligned. Instead, our approach uses a \textit{dynamic time warping} (DTW) kernel, which provides an elastic matching of two time sequences incorporating both the sequential ordering of the trajectory and the absolute values of time~\cite[]{Berndt1994}. The DTW similarity kernel is defined as:
\begin{equation}
\label{eq:DTW_sim}
k_l(x,y) = e^{-D_l(x,y)/\sigma^2},  
\end{equation}
where $D_l$ denotes the DTW distance. The distance $D_l$ is calculated by constructing an $n \times m$ matrix where $n$ and $m$ are the lengths of the two vectors we wish to compare. Using the pair-wise cost $cost(x_i,y_j) = ||x_i - y_i||^2 $, we minimise the overall cost over the path from $(i,j)=(1,1)$ to $(i,j)=(n,m)$ where each cell $(i,j)$ along the path contributes $cost(x_i,y_j)$ to the cumulative cost (summed over the path). This method is able to implicitly stretch both sequences to get a single dynamic time warping match between the two vectors, i.e., we find the cost required to match the two time-series trajectories for each learner. The higher the cost, the higher distance in Hilbert space, and therefore the lower similarity between learners. 

For $N$ learners we produce an $N \times N$ similarity matrix $A$ where each element $A_{ij}$ is the DTW similarity~\ref{eq:DTW_sim} between learners $i$ and $j$. For longer time-series and for larger number of learners $N$, whereby the DTW calculations may become computationally expensive, dimensionality reduction methods can be implemented to improve the speed of similarity calculations~\cite[]{Keogh2006} or segmented dynamic time warping algorithms with comparable speeds to Euclidean distances can be used~\cite[]{Keogh2010}. 

\subsubsection*{Creating a similarity graph using RMST sparsfication}
The similarity matrix $A$ can be thought of as the adjacency matrix of a fully connected, weighted graph, where every learner is connected to every other learner in the network with a different strength given by their pairwise similarity.  The high redundancy present in this full similarity matrix both increases the computation time and reduces the effectiveness of many clustering algorithms. We therefore sparsify the similarity matrix to produce a similarity graph by reducing the number of edges present. To do this, we employ a pruning algorithm (the Relaxed Minimum Spanning Tree, or RMST), which is based on geometric graph heuristics that preserves edges based on both their strength and their relevance to long paths within the graph. RMST has been shown to balance the local and global structure of datasets and performs well under multiscale graph clustering methods~\cite[]{Beguerisse-Diaz2013, Beguerisse2014}. 
Supplementary Figure~4 shows that the community structure is relatively stable when the sparsification parameter of RMST is varied.

Visualisations and layouts of the similarity graphs for the different datasets were produced using Gephi with the Force Atlas setting \cite[]{Bastian2009}.

\subsection*{Finding clusters of learners using Markov Stability graph partitioning}

Community detection methods for graphs aim to partition the nodes of a graph into subgraphs (communities) that are well-connected within themselves and weakly connected to each other. There are multiple ways to define communities, and many methods and criteria to score the resulting partitions \cite[]{Fortunato2010}. Such methods are also related to graph partitioning problems.

Markov Stability (MS) is a generalised method for identifying communities in graphs at all scales. 
MS employs a random walk on the graph to define a time-dependent cost function that measures the probability that a random walker is contained within a subgraph over a time scale $t$. 
If the random walker becomes trapped in particular subgraphs over that particular timescale, this identifies a good partition. 
As the time scale of the Markov process increases, the method identifies larger subgraphs leading to coarser partitions. 
Hence MS has the ability to identify intrinsically relevant communities at all scales by using the dynamic scanning provided by the diffusive process. For a detailed description of the method see~\cite{Delvenne2010,Lambiotte2014}.

The random walk is governed by the $N \times N$ transition matrix $Q=D^{-1}A$, where $N$ is the number of nodes in the graph, $A$ is the adjacency matrix, and $D=\text{diag}(A \mathbf{1})$ is the degree matrix where $\mathbf{1}$ is a vector of ones. $Q$ defines the probability of the random walk transitioning from node $i$ to node $j$, as given by the discrete-time process: 
\begin{equation}\label{eq:continuousrandomwalk}
\mathbf{p}_{t+1} = \mathbf{p}_t \, Q,
\end{equation}
where $\mathbf{p}_t$ is a $1 \times N$ node vector describing the probability of the random walker to be at each node at time $t$.
An associated continuous-time diffusive process in terms of the graph combinatorial Laplacian $L=D-A$ has the time-dependent solution:  
\begin{equation}\label{solutionrandomwalk}
\mathbf{p}(t) = \mathbf{p}(0) \, e^{- t L}.
\end{equation}
The time $t$ is denoted the \textit{Markov time} and is distinct to any real time. Markov time can be understood as a dimensionless quantity related to the diffusive process, which acts as a resolution parameter in that it allows for the exploration of the graph at different scales: as the Markov time increases, the partitions become coarser.

A partition of the graph into $c$ communities is encoded into a $N \times c$ membership matrix establishing the correspondence between the nodes and the clusters:
\begin{equation} 
    H_{ic}=\left\{
        \begin{array}{ll} 1 &  \mbox{if node $i$ belongs to community $c$} \\ 
            0 &   \mbox{otherwise.} 
        \end{array}
    \right.  \label{eq:Hdef} 
\end{equation} 
The goodness of the partition encoded by $H$ at time $t$ under the dynamics governed by $L$ is defined in terms of the $c \times c$  \textit{block auto-covariance} matrix:
\begin{equation}\label{autocovariance}
R(t;H) = H^T(\Pi  e^{- t L} -  \boldsymbol{\pi}^T \boldsymbol{\pi})H,
\end{equation}
where $\boldsymbol{\pi}$ is the stationary solution of~\ref{solutionrandomwalk} and $\Pi = \text{diag} (\boldsymbol{\pi})$. 
The meaning of this matrix is clear: the element of the matrix $[R(t; H)]_{\alpha \beta}$ encodes the probability that a random walker starting in community 
$\alpha$ will be at community $\beta$ after time $t$, and the diagonal elements, $[R(t,H)]_{\alpha \alpha}$, indicate the probability of remaining contained in community $\alpha$ over time scale $t$. Hence
a good partition $H$ will maximise the sum of the diagonal elements, i.e., the trace of $R(t,H)$. 
This leads us to our definition of the cost function, the Markov Stability of the partition:
\begin{equation}\label{markovstability}
r(t,H) = \min_{\tau < t} \text{Tr} \left[R(\tau,H)\right],
\end{equation}
which is to be maximised at every time $t$ by searching in the space of partitions $H$:
\begin{equation}\label{markovstability_opt}
r^*(t) = \max_{H} r(t,H)  \quad \text{ and } \quad H^*(t) = \argmax_{H} {r(t,H)}.
\end{equation}

Due to the optimisation~\ref{markovstability_opt} being non-convex and NP-hard, we use an efficient greedy algorithm known as the \textit{Louvain algorithm} \cite{Blondel2008}, which has been shown to perform well in practice and against benchmarks. Given its greedy nature, the optimised partition found by Louvain is not always the same as it depends on the initialisation of the optimisation algorithm. Therefore, we repeat the optimisation $\ell=100$ times using different starting points for the algorithm. For each Markov time we thus obtainn 100 optimised partitions $H_i^*(t)$ and we pick the one with maximal Markov Stability~\ref{markovstability} in the set as the optimal partition at $t$:
$$ \max_i \, \{H_i^*(t) \}_{i=1}^{\ell} = \widehat{H}(t).$$

To identify the important partitions across time, we use the following
two robustness criteria~\cite[]{Lambiotte2014}:
\paragraph{\textbf{Consistency of the optimised partition:}}
A relevant partition should be a robust outcome of the optimisation,
i.e., the ensemble of $\ell$ optimised solutions should be similar.
To assess this consistency, we employ an information-theoretical
distance between partitions: the normalised variation of information
between two partitions $\mathcal P$ and $\mathcal P'$ defined
as~\cite[]{Meila2003}:
\begin{equation}
\label{eq:VI}
 \textit{VI} (H, H') = \dfrac{2 \, \Omega(H,H') - \Omega(H) - \Omega(H')}{\log(N)},
\end{equation}
where $\Omega(H) = -\sum_{\mathcal C} p(\mathcal C) \log
p(\mathcal C)$ is a Shannon entropy, with $p(\mathcal C)$ given by the
relative frequency of finding a node in community $\mathcal C$ in the
partition $H$, and $\Omega(H, H')$ is the
Shannon entropy of the joint probability. 
The variation of information $\VI(H, H') \in [0,1]$ 
is a true metric distance between two partitions based on information theory 
and $\VI(H, H')=0$ indicates that two partitions are identical. 

A measure of the robustness to the optimisation, at a given Markov time $t$, 
is given by the average variation of information of the ensemble of solutions obtained from
the $\ell$ Louvain runs:
\begin{align}
\label{eq:VIt}
   \VI(t) = \dfrac{1}{\ell (\ell-1)} \sum_{i\neq j} \VI (H^*_i(t),H^*_j(t)).
\end{align} 
If all runs of the optimisation return similar partitions, then
$\VI(t)$ is small, indicating robustness of the partition to the
optimisation.  Hence we select partitions with low values (or dips) of
$\VI(t)$.

\paragraph{\textbf{Persistence of the partition across levels of resolution:}}
Relevant partitions should also be optimal across stretches of Markov
time.  Such persistence is indicated both by a plateau in the number of
communities over $t$ and a low value plateau of the cross-time
variation of information:
\begin{align}
\label{eq:VItt}
    \text{\VItt} = \VI (\widehat{H}(t),\widehat{H}(t')).
\end{align}  
This provides a second measure of robustness of a partition across resolution scales, and is commonly visualised via a heatmap where blocks along the diagonal indicate partitions that are persistent.
Within a time-block of persistent partitions we 
choose the most robust partition, i.e., with lowest $\VI(t)$.  

Markov Stability code available
at \url{github.com/michaelschaub/PartitionStability}.
When the computation of the matrix exponential in~\ref{autocovariance} becomes costly for moderately large $N$, the linearisation of $e^{- t L}$ provides an efficient approximate method to analyse very large graphs within the same framework.

\subsection*{Isotonic regression}

An isotonic regression is a model that identifies the optimal least squares fit to a data set given the constraint that the model must be a non-decreasing function. The optimisation is: 
\begin{equation}\label{isotonic}
\argmin_{x}\left | y-x \right |^{2} ,
\end{equation}
where $x_i$ must be larger or the same as $x_{i-1}$ i.e. $ x_0 \leq x_1 \leq ... \leq x_n$. The algorithm looks for violations of monotonicity and adjusts the estimate to fit within the constraints.

\subsection*{Gaussian Process Regression}

The Gaussian process regression (GPR) was implemented using the sklearn Python package. The implementation is based on the Algorithm 2.1 of Gaussian Processes for Machine Learning (GPML) by Rasmussen and Williams \cite[]{Rasmussen2010}.

A GPR model can be thought to define a distribution over functions and inference being undertaken directly on the space of functions. As such, a mean and variance that models the data can be calculated.
Given that the GPR is probabilistic we can calculate the log-likelihood of any set of trajectories being derived from an optimised GPR on another set of trajectories. Bayes factors are a method of Bayesian model comparison which quantify the support for a model over another model. The Bayes factor $K$ for two models $M_1$ and $M_2$ given some data $D$ is:
\begin{equation}\label{bayesfactor}
K = \frac{Pr(M_1 \mid D)}{Pr(M_2 \mid D)} \frac{Pr(M_2)}{Pr(M_1)}
\end{equation}

\subsection*{Additional classification algorithms}

To classify learners into high, medium and low performance groups, we used an SVM and a Decision Tree. Both algorithms are commonly used in classification tasks and were implemented using the scikit learn Python package \cite[]{PedregosaFabian2011}. 
\begin{itemize}
  \item An SVM acts as a non-probabilistic binary linear classifier that attempts to find a hyperplane in a high or infinite dimensional space that maximises the distances between data points of differing classes. We implemented the SVM with the radial basis function kernel.
  \item The Decision Tree attempts to find optimal branches (decisions) that represent conjunctions of features that lead to accurate prediction of class labels. We implemented a Decision Tree depth of 4 branches, increasing the number of branches did not improve the classification accuracy.
\end{itemize} 

Instead of using regression analysis between continuous dependent variables (performance) and independent variables (temporal features), we implemented classification algorithms to provide a closer comparison to our clustering results.

\subsection*{Code availability}

We have provided links to the necessary functions required for the mathematical framework detailed in this manuscript:

\begin{itemize}
  \item Clustering algorithm (Markov Stability) : \\
   \url{https://wwwf.imperial.ac.uk/~mpbara/Partition_Stability/} \\
   \url{https://github.com/michaelschaub/PartitionStability}
  \item Dynamics time warping: \\
   \url{https://github.com/pierre-rouanet/dtw}
\end{itemize}

\subsection*{Data availability}

To maintain anonymity of the learners that took part in this study we have not released the data.

\section*{Acknowledgements}

We would like to thank Nai Li, Marc Wells, Gavin Symonds, Samuel McGarry and Phil Tulip for assistance with data collection and interpretation. We would also like to thank Prof Alan Spivey for helping promote the project and attain funding from Imperial College London. We would like to thank Dr Iro Ntonia and Prof Martyn Kingsbury for their insightful suggestions and advice on ethical procedures.

This research has been funded by a President's Excellence Award from Imperial College London. 
M.B. and S.N.Y. acknowledge support from EPSRC award EP/N014529/1 funding the EPSRC Centre for Mathematics of Precision Healthcare at Imperial.

\section*{Competing Interests}

The authors have no competing interests to declare.

\section*{Author Contributions}

R.L.P. and M.B. designed the mathematical framework and data analytics pipeline. R.L.P. built and coded the Python toolbox for data collection, data cleaning and data analysis, and carried out the computational analysis. M.B, S.N.Y and D.L supervised the project design and data analytics research. All authors contributed to writing the manuscript.

%
\bibliographystyle{unsrt}


\end{document}